\documentstyle[12pt]{article}
\def\d{\partial}

\def\pmb#1{\setbox0=\hbox{#1}%
\kern.0em\copy0\kern-\wd0
\kern-.04em\copy0\kern-\wd0
\kern.08em\copy0\kern-\wd0
\kern-.04em\raise.0433em\box0 }         

\newcommand{\nc}{\newcommand}
\nc{\pek}[1]{\cite{#1}}
\nc{\enr}[1]{(\ref{#1})}
\nc{\kal}[1]{{\cal{#1}}}

\def\bop#1{\setbox0=\hbox{$#1M$}\mkern1.5mu
        \vbox{\hrule height0pt depth.04\ht0
        \hbox{\vrule width.04\ht0 height.9\ht0 \kern.9\ht0
        \vrule width.04\ht0}\hrule height.04\ht0}\mkern1.5mu}

\begin{document}

\newcommand{\inv}[1]{{#1}^{-1}} 

\renewcommand{\theequation}{\thesection.\arabic{equation}}
\newcommand{\beq}{\begin{equation}}
\newcommand{\eeq}[1]{\label{#1}\end{equation}}
\newcommand{\ber}{\begin{eqnarray}}
\newcommand{\eer}[1]{\label{#1}\end{eqnarray}}
\begin{center}

                                \hfill    USITP-00-04 \\
                                \hfill    hep-th/0005142\\

\vskip .3in \noindent

\vskip .1in

{\large \bf {Hamiltonian systems with boundaries}}
\vskip .2in

 {\bf Maxim Zabzine\footnote{zabzin@physto.se}}\\

\vskip .15in

\vskip .15in

\vskip .15in
{\em  Institute of Theoretical Physics,
University of Stockholm \\
Box 6730,
S-113 85 Stockholm SWEDEN}\\
\bigskip

\vskip .15in

\vskip .1in
\end{center}
\vskip .4in
\begin{center} {\bf ABSTRACT } 
\end{center}
\begin{quotation}\noindent 
 Lately,  to provide a solid ground for quantization of the
 open string theory with a constant $B$-field, it has been proposed 
 to treat the boundary conditions as  hamiltonian constraints.
  It seems that this proposal is quite general and  should be
 applicable to a wide range of models defined on manifolds with 
 boundaries. The goal of the present paper is to show
 how the boundary conditions can arise  as constraints 
 in a purely algebraic fashion within the Hamiltonian 
 approach without any reference to the Lagrangian
 formulation of the theory. The construction of the
 boundary Dirac brackets is also given and some subtleties are pointed out.  
  We consider four examples of field 
 theories with boundaries: the topological sigma model,
 the open string theory with and without a constant 
 $B$-field and electrodynamics with topological term.     
  A curious result for electrodynamics on a manifold
  with boundaries is presented.

\end{quotation}
\vfill
\eject

\section{Introduction and motivation}

  Lately, there has been a renewed interest in  field theories on  
 manifolds with  boundaries. In general one would expect 
 a nontrivial relation between bulk and boundary dynamics.
 One such model is the 
 open string theory with a constant $B$-field. It turns out that the bulk
 and boundary properties of this model are quite different \cite{Chu:1999qz}.
  Recently, in the attempt to provide a 
 solid ground for the quantization of the 
 model, it has been proposed to treat  boundary conditions  as 
 Hamiltonian constraints within the Dirac approach \cite{Ardalan:1999av,
 Chu:1999gi, Sheikh-Jabbari:1999xd, Kim:1999qs, Lee:1999kj}. 
 This idea seems quite 
 powerful and could be applied to a wide range of models. 

 Let us recall that for systems with boundaries, traditionally
 the boundary
 conditions have to be imposed to  properly define the
 functional derivatives in the theory. Specifically, for the 
 Hamiltonian treatment one needs a boundary conditions for
 the proper definition of the Poisson brackets (symplectic structure
 on the phase space). However, there is an alternative algebraic approach 
 to the definition of symplectic structures.
 Using three basic properties of the Poisson bracket (antisymmetry,
 Leibniz rule and Jacobi identity) and the canonical brackets for
  momenta and coordinates one can calculate any bracket. 
 Of course a function on the phase space
 should now be understood as a formal power expansion in  
 momenta and coordinates. This approach to the Poisson bracket is
 in the spirit of quantum mechanics where the algebraic definitions are
 the basic ones. 
  
 Thus applying the algebraic definition of Poisson bracket one 
 does not have to impose  the boundary conditions in 
 order to do concrete calculations.  
 As a matter of fact one sees that  to define 
 momentum, hamiltonian and primary constraints formally 
 in many models there is no need to use the
  boundary conditions. Thus the natural question arises:
 what is the status of boundary conditions in this framework. 
 In this paper we try to answer this question. 
  The main point which we are going to make is that the boundary 
 conditions can arise in a purely algebraic fashion as
  Hamiltonian constraints localized on the boundary. As a result 
 all the Dirac machinery may be applied\footnote{That is true
 up to certain technical problems which, 
 we believe, can be resolved.}  to the boundary conditions 
 vs the Hamiltonian boundary constraints. Indeed we need 
 those boundary constraints
 to make the whole Hamiltonian treatment consistent.

 Let us make a few technical remarks. 
 The basic idea is rather naive.
 Since one can formally define the  momentum, hamiltonian,
 the primary constraints and do calculations with the Poisson brackets 
  without using  the boundary conditions
 we may proceed in a formally  along Dirac's lines using the 
 canonical Poisson brackets \cite{Dirac}. However now 
 we are not allowed 
 to throw away the total derivative terms (the boundary terms).
 These boundary terms produce the corresponding constraints 
 on the boundary. The usual consistency conditions have to be required
 for these constraints. To handle the technical side of the
 idea it is useful to work with constraints $\Phi$ smeared with test 
 functions $N(x)$
\beq
 \Phi[N] = \int d^dx \,\Phi(x)\, N(x).
\eeq{0.0} 
 This is often a convenient notation, especially when one wants
 to keep track of partial integrations in a calculation. For the case
 of boundary constraint one assumes that the smearing function
 has support on the boundary only. (Those assumptions do
 not effect the formal calculations in any way.) In all calculations we will
 avoid the functional questions and will concentrate attention on
 the algebraic aspect of the computations. We will see that there 
 is no ambiguity as soon as a calculation is done using  
 test functions. However to define the Dirac brackets 
 one has to do calculations without the test functions (as far
 as the author knows) and this may lead to trouble in some cases. 
 To make sense of those boundary Dirac's bracket 
 in certain cases one should give 
 a mathematically rigorous definition of the relevant objects.
 We do not do this and just present the formal answer
 with short comments.   
 We clarify this point by considering concrete examples. 
   
 It is worthwhile to make some remarks concerning the status of
  boundary conditions also within the Lagrangian formalism. The presence
  of a boundary can spoil some properties which hold in the 
  situation  without a boundary.
  For instance, two classically equivalent actions
 (in the sense that they reproduce the same equations of motion)
 can give rise to different boundary conditions. A nontrivial 
 example of this situation is the relation between the Howe-Tucker and
 Nambu-Goto actions for the open branes. These actions give slightly
  different boundary conditions.
 To relate them is a subtle task and  
 depends on the  dimension of  the background 
 space-time\footnote{For instance, the strings in
 two dimensional space-time: The Nambu-Goto action is linear 
 and it gives rise only the Dirichlet boundary conditions. However
 the Howe-Tucker (Polyakov) action is quadratic and 
 it might produce as well the Neumann boundary conditions.}.
 This is another reason for looking at the Hamiltonian treatment of the 
 boundary conditions. 

 The main motivation behind the present work is  to
 understand the status of the boundary conditions in a quantum
 theory especially an interacting one.  
 The Hamiltonian approach has certain advantages when it
 comes to quantizing a theory. (At least in principle it
 is clear what one should do.) At the end we will comment on 
 the possible  quantum applications of our results. 

 Let us briefly comment on the literature. In mathematical physics 
 the Hamiltonian systems with boundaries is an old subject (see 
 \cite{Solovev:1993zf} for a list of references).
 The main attention has been on possible 
 modifications of the Poisson bracket by surface terms 
  to fulfil general axiomatic 
 properties in the presence of boundaries. A modifed Poisson bracket 
 was defined in \cite{Solovev:1993zf} and 
  later generalized in \cite{Bering:1998fm}.   
  Unlike \cite{Solovev:1993zf, Bering:1998fm} the present discussion
 is not  formal. Our attitude is conservative. We calculate
 the  Poisson brackets in the standard way and keep track of the
 boundary terms which we interpret as Hamiltonian constraints. 
 Eventually these Hamiltonian boundary constraints should lead
 to a Dirac bracket which gives the right symplectic 
 structure for the model.  

 Since the general idea by itself is simple one and it is 
 difficult to give any general theorems,  examples are
 quite helpful. Thus throughout the paper we consider four examples:
 the topological sigma model with a boundary, 
 the open string theory with and without a constant $B$-field and four 
 dimensional $U(1)$ gauge theory on a manifold with a boundary. 
 There is a section for every  example and  at the end 
 we summarize the results and discuss the problems.
 In the first section we consider the topological sigma model
  with boundaries. This example is rather simple and particular. 
 It demonstrates that there is a difference between the functional and 
 algebraic approaches to the Poisson bracket and that the former approach
 misses some interesting information about the boundary.  
  In next two section we consider the open string theory.
 We show that the boundary constraints can arise in a purely
 algebraic fashion from the algebra of constraints. We hope that
 the discussion in the fourth section will clarify some points 
 in earlier analysises of the problem \cite{Ardalan:1999av,
 Chu:1999gi, Sheikh-Jabbari:1999xd, Kim:1999qs, Lee:1999kj}. We also point
 out that some problems may arise in the definition of modified
 symplectic structure on the boundary.  
  The last section is devoted to Euclidean electrodynamics 
 with a topological term on a manifold with boundaries. 
 In the spirit of open string theory with B-field we
 derive the modification of the symplectic structure on 
 the boundary.

\section{Topological sigma model with boundaries}

 Let us consider a topological sigma model with 
 boundaries defined on 2d-dimensional smooth manifold which 
 admits a symplectic 
 structure\footnote{$\omega=\omega_{\mu\nu}dX^\mu\wedge dX^\nu$ 
 is a symplectic structure if $d\omega =0$ and $\omega_{\mu\nu}$ is
 not degenerate.} $\omega$ 
\beq
S= \frac{1}{2}\int\limits_{\Sigma} 
 d^2\xi\, \omega_{\mu\nu}(X) \partial_\alpha X^\mu 
 \partial_\beta X^\nu \epsilon^{\alpha\beta} ,
\eeq{a.1}
 where $\Sigma$ is a two-dimensional world-sheet
 with boundary. In the bulk this model is purely topological and
 has no local degrees of freedom \cite{Witten:1988xj}.
  For the present purposes
 we ignore the topological aspects of the model and thus 
 assume that the background space-time manifold can be covered
 by one patch. It means that we can think of the sympelctic form
 as an exact two form $\omega = dA$ and the action (\ref{a.1}) becomes
\beq
 S = \int\limits_{\partial \Sigma} d\tau A_\mu(X) \dot{X}^\mu , 
\eeq{a.2} 
 which describes the boundary dynamics. These boundary dynamics are
  trivial ($\dot{X}^\mu=0$) and there is just a modification
 of the Poisson brackets for $X^\mu$ since there are 2d second class
 constraints.  

 Now we want to try to extract the information about boundary 
 dynamics from the Hamiltonian treatment, starting from the 
 action (\ref{a.1}).
A variation of the action (\ref{a.1}) gives
\beq
 \delta S =  \int\limits_{\d\Sigma} d\xi^\alpha 
 \omega_{\mu\nu} \delta X^\mu \partial_\alpha X^\nu +
 \frac{1}{2} \int\limits_{\Sigma} 
 d^2\xi\, (d\omega)_{\mu\nu\rho} \partial_\alpha X^\mu 
 \partial_\beta X^\nu \epsilon^{\alpha\beta} \delta X^\rho , 
\eeq{a.3}  
 where the last term vanishes by itself (see footnote) 
 and the boundary term should vanish as well.
 Since $\omega_{\mu\nu}$ is nondegenerate, one should impose
 the Dirichlet boundary condition
\beq
 \delta X^\mu |_{\d\Sigma} = 0, 
\eeq{a.4}
  which simply means that the naive functional derivative with respect to 
 $X$ is not defined on the boundary. Thus the functional approach
 cannot be used to find the boundary dynamics.
 Instead we may use an algebraic approach to the problem.
 The action (\ref{a.1}) produces
 $2d$ constraints 
\beq
 \Phi_\mu[N^\mu] = \int d\sigma N^\mu 
 (P_{\mu} -  \omega_{\mu\nu}(X) X'^\nu) , 
\eeq{a.5}
 which give the following Poisson bracket algebra
\beq
 \{ \Phi_\mu[N^\mu], \Phi_\nu[M^\nu] \} =
  \int d\sigma \left [(d\omega)_{\mu\nu\rho} N^\mu M^\nu X'^\rho \right ]-
 N^\mu M^\nu \omega_{\mu\nu} |_{0}^{\pi},  
\eeq{a.6} 
 where $N^\mu$ and $M^\nu$ are test functions and $\sigma \in [0,\pi]$.
 In the present calculation (and as well as in the next two sections) 
 we use the following formula
\beq
 \int\limits_{0}^\pi d\sigma  \int\limits_{0}^\pi d\sigma'
 f(\sigma) g(\sigma') \partial_\sigma(\delta(\sigma- \sigma')) =
 f(\sigma)g(\sigma)|_{0}^{\pi} - \int\limits_{0}^\pi d\sigma f'(\sigma)
 g(\sigma).
\eeq{a.6a}
 which can be easily motivated.
  The constraints (\ref{a.5}) are first class in the bulk and
 second class constraints on the boundary. To simplify the calculations
 we can do the following.
 Since we are working on 
 one patch one can assume that the symplectic 
 structure $\omega_{\mu\nu}$ is a constant matrix of a special form 
 (due to the Darboux theorem in one patch 
 there are always special coordinates where the symplectic form can 
 be brought to  canonical form). 
 Thus because of (\ref{a.6}) there is
 a suitable modification of the symplectic structure on the boundary
\beq
 \{ X^\mu, X^\nu\}|_{\d\Sigma} = \omega^{\mu\nu} ,
\eeq{a.7}
 where $\omega^{\mu\nu} \omega_{\nu\rho} = \delta_\rho^\mu$.
 Furthermore,
\beq
 \{ P_\mu, P_\nu\}|_{\d\Sigma} = -\frac{1}{4}\omega_{\mu\nu},
\,\,\,\,\,\,\,\,\,\,\,
  \{ X^\mu, P_\nu\}|_{\d\Sigma} = \frac{1}{2} \delta_\nu^\mu,
\,\,\,\,\,\,\,\,\,\,\, 
\{X'^\mu,  X^\nu\}|_{\d\Sigma} = \frac{1}{2} \omega^{\mu\nu}.
\eeq{a.8} 
 It is easy to check that all these brackets have the 
 desired properties (everything should have a trivial bracket 
 with constraints on the boundary).  
 Proceeding along standard lines one finds that $\delta X^\mu = 
 \dot{X}^\mu$ equals $N^\mu$ in the bulk and zero on the boundary. 
  
 The present model is  trivial, nevertheless it contains the essence
  of the general situation of Hamiltonian models with
 boundaries. It shows that there is a difference between the functional 
 and the algebraic approaches to the symplectic structure. In the algebraic
 approach the right boundary conditions arise by themselves in a consistent
 way. In the next sections we consider less trivial examples of this
 situation. 

\section{Open string theory without a $B$-field}

 Let us consider an open string theory in a flat space-time
 ($\eta_{\mu\nu}=(-1,1,...,1)$) without antisymmetric background field. 
 The model has the following action
\beq
 S= -\frac{1}{2} \int d^2\sigma\,\,\sqrt{-h} h^{\alpha\beta} 
 \d_{\alpha} X^\mu \d_\beta X^\nu \eta_{\mu\nu},
\eeq{1.0}
 where $h^{\alpha\beta}$ is an auxilary metric. The treatment
 of the theory is presented in string theory textbooks  (for
 instance \cite{Green:1987sp}). To the author's knowledge the 
 canonical treatment of the open string  has only been given
 in the lectures  by Henneaux \cite{Brink:1988nh}. In this section we  
 would like to have a new look  at some 
 well-known facts about open strings.
 The action (\ref{1.0}) produces the following boundary condition
\beq
 (\sqrt{-h} h^{10} \dot{X}^\mu + \sqrt{-h} h^{11} X'^\mu)|_{0,\pi}=0 .
\eeq{1.0a}
 If one starts from the Nambu-Goto action instead then the general 
 boundary condition is  
  $\dot{X}^\mu \sim  X'^\mu|_{0,\pi}$ which is equivalent to (\ref{1.0a}). 
 The condition
 (\ref{1.0a}) can be rewritten as follows in phase space 
\beq
(\eta_{\mu\nu} X'^\nu +  \sqrt{-h} h^{01} P_\mu)|_{0,\pi}=0 ,
\eeq{1.0b} 
 which states that $\eta_{\mu\nu} X'^\nu$ and $P_\mu$ are proportional
 to each other on the boundary.

 Now let us turn to the Hamiltonian analysis of the system.
 For the model (\ref{1.0}) the constraints are well known
\beq
{\cal H}_1 [N] = \int\limits_{0}^{\pi} d\sigma P_\mu X'^\mu N,
\,\,\,\,\,\,\,\,\,\,\,\,\,\,\,
 {\cal H} [M] = \int\limits_{0}^{\pi} d\sigma (P_\mu \eta^{\mu\nu} P_\nu +
 X'^\mu \eta_{\mu\nu} X'^\nu) M,
\eeq{1.1}
 and they hold at all points including the boundary.  
 Since the system is generally covariant the naive Hamiltonian
 vanishes identically. Both constraints (\ref{1.1}) are first
 class  and they correspond to reparametrizations of the
 two dimensional world sheet. 
 The constraints obey the following Poisson bracket algebra
\ber
 \{ {\cal H}_1[N], {\cal H}_1[M]\} = {\cal H}_1[NM'-N'M],\\
 \{ {\cal H}_1[N], {\cal H}[M]\} = {\cal H}[NM'-N'M] + 
 NM (P_\mu \eta^{\mu\nu}P_{\nu} - X'^\mu \eta_{\mu\nu} X'^\nu)|_{0}^{\pi},\\
 \{ {\cal H}[N], {\cal H}[M] \} = {\cal H}_1[4(NM'-N'M)].
\eer{1.2}
 The bracket between ${\cal H}_1$ and ${\cal H}$ gives rise the
 boundary term which should be set to zero to make the Hamiltonian
 treatment consistent. Since ${\cal H}_1$ and ${\cal H}$ hold
 everywhere we must require the following constraints
 on the boundary
\beq
P_\mu \eta^{\mu\nu}P_{\nu}|_{0,\pi} =0, \,\,\,\,\,\,\,
\,\,\,  X'^\mu \eta_{\mu\nu} X'^\nu|_{0,\pi} =0,\,\,\,\,\,\,\,
\,\,\,P_\mu X'^\mu|_{0,\pi} =0.
\eeq{1.3} 
 One might call them the boundary constraints. The next step should be
 to check whether the algebra of new constraints is closed or not. 
 As we said before all calculations can be done in
 a formal way avoiding questions of regularization. For example
 let us introduce the following notation for the boundary constraints
\beq
 \phi_1[N] = \int\limits_{0}^{\pi} d\sigma\, N\,P_\mu \eta^{\mu\nu}P_{\nu},
\,\,\,\,\,\,\,\,\,\,\,\,
 \phi_2[M] =  \int\limits_{0}^{\pi} d\sigma\,M\,X'^\mu \eta_{\mu\nu} X'^\nu
\eeq{1.4}
 where $N$ and $M$ might be thought as test functions localized
 on the boundary (or around boundary if there is some 
 regularization assumed). This kind 
 of assumptions does not effect the formal calculations. For instance
 we calculate the following brackets
\beq
 \{ \phi_1[N], \phi_2[M] \} = 4\int\limits_{0}^{\pi} d\sigma\, NM'
 P_\mu X'^\mu +  4\int\limits_{0}^{\pi} d\sigma\, NM
 P_\mu X''^\mu  - 4NMP_\mu X'^\mu|_{0}^{\pi}, 
\eeq{1.5}
 and see that secondary constraints arise. However the 
 constraints (\ref{1.3}) can be resolved since $P$ and 
 $X'$ are null vectors on the boundary and
 they are orthogonal to each other there is a proportionality
 relation on the boundary
\beq
 (\alpha P_\mu + \eta_{\mu\nu} X'^\nu)_{0,\pi} =0,
\eeq{1.6} 
 where $\alpha$ is some proportionality constant which is subject
 to gauge condition (since it relates world-sheet density to the
 world-sheet vector). The conditions (\ref{1.6}) give us 
 the same information as 
 one would get from the Lagrangian formalism (\ref{1.0b}).
 Hence the whole system can be described as
 two first class constraints ${\cal H}_1$, ${\cal H}$ plus a set
 of second class boundary constraints (\ref{1.6}). The constraints 
 (\ref{1.6}) are second class because of the non vanishing brackets
\beq
 \{ \Phi_\mu[N^\mu], \Phi_\nu[M^\nu]\} = 
 \alpha \int\limits_{0}^{\pi} d\sigma\,[N^\nu M'^\mu -
 N'^\mu M^\nu] \eta_{\mu\nu},
\eeq{1.7}   
 where $\Phi_\mu[N^\mu]$ is (\ref{1.6}) smeared with the test function $N^\mu$.
 Proceeding formally for the second class constraints (\ref{1.6})
 we define the corresponding Dirac brackets
\beq
 \{ X^\mu(\sigma), X^\nu(\sigma')\} = \frac{\alpha}{2} \eta^{\mu\nu}
 \frac{1}{\partial_\sigma} \delta(\sigma-\sigma') ,
\eeq{1.8} 
 as well as the brackets
\beq
 \{ X^\mu(\sigma), P_\nu(\sigma')\} = \frac{1}{2} \delta^\mu_\nu 
 \delta(\sigma-\sigma'),\,\,\,\,\,\,\,\,\,\,
 \{ X'^\mu(\sigma), X^\nu(\sigma')\} = \frac{\alpha}{2} \eta^{\mu\nu}
  \delta(\sigma-\sigma').
\eeq{1.9}
 We are interested in the restriction 
 of these brackets to the boundary and it is not clear how to find this,
 especially for the non-local bracket (\ref{1.8}). The point is that this
 question can not be answered unless our  description of
 the model is supplemented with a certain amount of additional information.
 The extra information concerns the restrictions on the behaviour
 of the fields in order to make operator $\partial_\sigma$ invertable
 (in general there is a constant zero mode for this operator). Therefore
 to make further progress one needs more insight into the model.
 It would be interesting to quantize the free open string theory in a
 nonconformal gauge (where $\alpha\neq 0$) and calculate the commutators
 (\ref{1.8}), (\ref{1.9}) explicitly on the boundary.   
  Resolving this kind of questions can lead to the proper understanding 
 of the foundations of  Witten's open string field 
 theory \cite{Witten:1986cc} where the noncommutativity of the ends of 
 strings plays a crucial role.

\section{Open string theory with a constant $B$-field}

 Now let us turn to the open string theory
 with a constant $B$-field. The model has the following action
\beq
S= - \frac{1}{2} \int d^2\sigma\,(\sqrt{-h} h^{\alpha\beta} 
 \d_{\alpha} X^\mu \d_\beta X^\nu \eta_{\mu\nu} - \epsilon^{\alpha\beta}
 \d_{\alpha} X^\mu \d_\beta X^\nu B_{\mu\nu}).
\eeq{2.0}
 This system has attracted much attention recently because of the 
 noncommutative properties of the end points of the string. A treatment
 of the model has been given in \cite{Chu:1999qz} (also 
 see \cite{Seiberg:1999vs}
 for the quite full list of references). Let us just  recall 
 that in the Lagrangian formalism one should impose the 
 boundary conditions 
\beq
(\sqrt{-h} h^{10}\eta_{\mu\nu} \dot{X}^\nu + 
 \sqrt{-h} h^{11}\eta_{\mu\nu} X'^\nu
 + B_{\mu\nu} \dot{X}^\nu) |_{0,\pi} =0,
\eeq{2.0a}
 which have the following form in phase space
\beq
 (B_\mu{^\nu} P_\nu + G_{\mu\nu} X'^\nu + \sqrt{-h}h^{01}P_\mu)|_{0,\pi} =0,
\eeq{2.0b}
 where $G_{\mu\nu} = \eta_{\mu\nu} - B_{\mu\sigma} B^{\sigma}{_\nu}$.
 For the sake of simplicity we assume that $B$ is a non-degenerate
 matrix (for the degenerate case one can easily generalize all the following 
 arguments).

 Now we turn to the Hamiltonian formalism.
 In the usual fashion the constraints are
\beq
{\cal H}_1 [N] = \int\limits_{0}^{\pi} d\sigma P_\mu X'^\mu N,
\eeq{2.1a}
\beq
 {\cal H} [M] = \int\limits_{0}^{\pi} d\sigma (P_\mu \eta^{\mu\nu} P_\nu -
 2 P_\mu B^{\mu}{_\nu} X'^\nu 
+ X'^\mu G_{\mu\nu} X'^\nu) M.
\eeq{2.1b}
 These are first class constraints and they 
  hold everywhere including at the boundary points. Next we
 calculate the algebra keeping track of the boundary terms. 
  The constraints obey the following Poisson bracket algebra
\ber
 \{ {\cal H}_1[N], {\cal H}_1[M]\} = {\cal H}_1[NM'-N'M],\label{2.2a}\\
 \{ {\cal H}_1[N], {\cal H}[M]\} = {\cal H}[NM'-N'M] + 
 NM (P_\mu \eta^{\mu\nu}P_{\nu} - 
 X'^\mu G_{\mu\nu} X'^\nu)|_{0}^{\pi},\label{111}\\
 \{ {\cal H}[N], {\cal H}[M] \} = {\cal H}_1[4(NM'-N'M)]. 
\eer{2.2}
 To make the theory consistient one should set the boundary term to zero.
 Since ${\cal H}_1$ and ${\cal H}$ hold everywhere 
 there is a boundary constraint 
\beq
  X'^\mu (B_\mu{^\nu} P_\nu + G_{\mu\nu} X'^\nu) |_{0,\pi} =0 ,
\eeq{2.3}
 which is the difference between ${\cal H}_1$ and the boundary term in 
 (\ref{111}). One cannot solve the system as simply as before.
 Therefore we proceed along Dirac's lines
 \cite{Dirac}. We look at possible secondary and tertiary 
 constraints and then try to separate them into first and second class
 constraints. Sometimes, before separating them 
 into different classes it is helpful to solve some of them. 
 
 We thus have to calculate brackets 
 of all constraints including the boundary one and see if 
 new constraints arise. We will perform the calculations in a formal way and 
 introduce the following notation for the boundary constraint
\beq
 \Phi[N] = \int\limits_{0}^\pi d\sigma\, N\, 
 X'^\mu (B_\mu{^\nu} P_\nu + G_{\mu\nu} X'^\nu),
\eeq{2.4a} 
 where $N$ is a test function. As a result of the computations some new 
 constraints  will arise. Let us look at some of them  
 to see the pattern. We have
\beq
 \{ \Phi[N], \Phi[M]\} = \int\limits_{0}^{\pi} d\sigma\,[NM'-N'M]
 X'^\mu B_\mu{^\rho}(B_\rho{^\nu} P_\nu + G_{\rho\nu} X'^\nu). 
\eeq{2.4}
 Introducing the following notation for the new constraint 
\beq
\Phi_1[N]  = \int\limits_{0}^{\pi} d\sigma\,N  
 X'^\mu B_\mu{^\rho}(B_\rho{^\nu} P_\nu + G_{\rho\nu} X'^\nu),
\eeq{2.4b}
 we get 
\beq
\{ \Phi_1[N], \Phi_1[M]\} = \int\limits_{0}^{\pi} d\sigma\,[NM'-N'M]
 X'^\mu B_\mu{^\delta} B_\delta{^\sigma} B_\sigma{^\rho} 
 (B_\rho{^\nu} P_\nu + G_{\rho\nu} X'^\nu),
\eeq{2.5}
 and 
\beq
\{  \Phi_1[N], \Phi[M]\} = \int\limits_{0}^{\pi} d\sigma\,[NM'-N'M]
 X'^\mu B_\mu{^\delta} B_\delta{^\rho} 
 (B_\rho{^\nu} P_\nu + G_{\rho\nu} X'^\nu), 
\eeq{2.6}
 and so on. This suggests the following  boundary conditions 
\beq 
X'^\mu M_\mu{^\sigma}
 (B_\sigma{^\nu} P_\nu + G_{\sigma\nu} X'^\nu) |_{0,\pi} =0,
\eeq{2.6a} 
 where $M$ is some power of $B$. Since $B$ is nondegenerate and
 antisymmetric all these
 conditions can be replaced by the following one
\beq
 (B_\mu{^\nu} P_\nu + G_{\mu\nu} X'^\nu + \beta P_\mu) |_{0,\pi} =0,
\eeq{2.6b}
 where $\beta$ is the coefficient of proportionality which is 
 subject to a gauge condition (like $\alpha$ in the previous section).
We will see that (\ref{2.6b}) are second class constraints. 
 Introducing the notation  
\beq
 {\cal K}_\mu [N^\mu] =  \int\limits_{0}^{\pi}d\sigma\, N^\mu
 (B_\mu{^\nu} P_\nu + G_{\mu\nu} X'^\nu + \beta P_\mu)
\eeq{2.7}
 it is easy to check the brackets
\beq
\{ {\cal K}_\mu [N^\mu], {\cal K}_\nu [M^\nu]\} =
 \int\limits_{0}^{\pi} d\sigma\,[N^\mu M'^\nu - N'^\nu M^\mu] (B_\mu{^\rho}
 G_{\rho\nu} + \beta G_{\mu\nu})
\eeq{2.8}
 where we have nondegenerate matrix on the  right-hand side. 
 Therefore we conclude that to make the whole Hamiltonian 
 treatment consistent one must impose the boundary conditions (\ref{2.0b})
 which play the role of second class constraint on the boundary.
 Otherwise the algebra (\ref{2.2}) would not be closed. 
 Thus the bracket algebra has to be modified on the boundary.
 The Poisson bracket must be replaced by the Dirac bracket.
 For example on the boundary the coordinates have
 the following bracket
\beq
\{ X^\mu, X^\nu \}_{\partial\Sigma} = - 
 B^\mu{_\sigma} (G^{-1})^{\sigma\nu} + 
 \beta (nonlocal\,\,part)
\eeq{2.9}
 where the non local part has the same structure as in the previous 
 section. For the case $\beta =0$ (for instance, conformal gauge or
 static gauge) the brackets (\ref{2.9}) are well defined.
 A discussion of the modified brackets is given in  \cite{Ardalan:1999av,
 Chu:1999gi, Sheikh-Jabbari:1999xd, Kim:1999qs, Lee:1999kj}.

\section{Electrodynamics with topological term}

As a last example we consider  theory with
 a nonvanishing Hamiltonian. We will take a look at four dimensional 
 Euclidean electrodynamics with a topological term. 
   The action is defined by 
\beq
S= \frac{1}{2g^2} \int\limits_{\cal M} F\wedge *F 
 + \frac{i\theta}{4\pi^2} \int\limits_{\cal M} F \wedge F ,
\eeq{3.1}
 where we use differential forms. Equivalently, in
  components, 
\beq
 S = \frac{1}{4g^2} \int\limits_{\cal M} d^4x\, F_{\mu\nu}
 F^{\mu\nu} + \frac{i\theta}{16\pi^2} \int\limits_{\cal M} d^4x\,
 \epsilon_{\mu\nu\rho\sigma}F^{\mu\nu} F^{\rho\sigma} .
\eeq{3.1q}
 The theory
 is defined on a manifold ${\cal M}$ with non empty boundary 
 $\d {\cal M}$. Since we are interested in the Hamiltonian treatment 
 we assume that ${\cal M}= R \times \Sigma$ where $\Sigma$ is a spatial
 manifold. For the sake of simplicity we further assume that $\Sigma$ is
 closed set in $R^3$ and thus it carries a flat metric. This assumption is
 not essential  and the whole logic can be generalized to the 
 general curved case. 

 Before looking at the Hamiltonian formalism we briefly consider
 the Lagrangian formalism. To the author's knowledge this system
 has not been separately studied, except in \cite{Bordag:1999ux}. 
 The action (\ref{3.1}) 
 gives the following equations of motion 
\beq
 d*F=0,\,\,\,\,\,\,\,\,\,\,\,\,\,\,\,\,\,\,
\eeq{3.1a}
 which should be supplemented by the boundary condition
\beq
\int\limits_{\d {\cal M}} \delta A \wedge (\frac{1}{g^2}*F + 
 \frac{i\theta}{2\pi^2}F) = 0. 
\eeq{3.2}
 To proceed further let us write (\ref{3.2}) in components
\beq
-\int dt \int\limits_{\d \Sigma} d^2s\,\left [ n^a (\frac{1}{g^2}E_a +
\frac{i\theta}{2\pi^2}B_a)\delta A^0 - n_b \epsilon^{abc} \delta A_c
 (\frac{1}{g^2} B_a + \frac{i\theta}{2\pi^2} E_a)\right ],
\eeq{3.2a}
 where we introduce the standard 
 notation $E_a\equiv F_{0a}$ and $B_a = \frac{1}{2} 
 \epsilon_{abc}F^{bc}$ and $n^a$ is a vector normal to $\d \Sigma$.
 Now it is straightforward to read off
 the boundary conditions
\beq
 n^a (\frac{1}{g^2} E_a + 
 \frac{i\theta}{2\pi^2}B_a)|_{\d {\cal M}}=0\,
\,\,\,\,\,\,\,\,\,\,\,\,\,\,\,or\,\,\,\,
\,\,\,\,\,\,\,\,\,\,\,\,
 \delta A^0|_{\d {\cal M}} =0,
\eeq{3.2a}
\beq
n_b \epsilon^{abc} (\frac{1}{g^2} B_a + 
\frac{i\theta}{2\pi^2} E_a)|_{\d {\cal M}}=0\,
\,\,\,\,\,\,\,\,\,\,\,\,\,\,\,or\,\,\,\,
\,\,\,\,\,\,\,\,\,\,\,\,
 \delta A_c |_{\d {\cal M}} =0.
\eeq{3.2b}
 Thus one of these two sets of conditions should be imposed
 to make the Lagrangian treatment consistent\footnote{Apart from this
 one can think about other physical requirements such as  absence of
 energy-momentum flow through the boundary. In fact the conservation
 of energy-momentum requires  the conditions on the
 left hand side of (\ref{3.2a}) and (\ref{3.2b}) \cite{nuno}.}. 

  Let us rewrite the boundary conditions (\ref{3.2a}), (\ref{3.2b})
  in phase space. The momentum is defined as follows 
\beq
 \pi_a = \frac{1}{g^2} E_a + \frac{i\theta}{2\pi^2} B_a,
\eeq{3.3}
 and  there is the usual constraint $\pi_0=0$ which we will 
 discuss later on. Using (\ref{3.3})  
 the boundary conditions (\ref{3.2a}), (\ref{3.2b}) become 
\beq
 n^a \pi_a|_{\d \cal M} =0\,\,\,\,\,\,\,\,\,\,\,\,\,\,\,\,\,\, or
\,\,\,\,\,\,\,\,\,\,\,\,\,\,\,\,\,\,\delta A^0|_{\d \cal M} =0,
\eeq{3.4}
\beq
 n_b \epsilon^{abc}
 \left ( \frac{i\theta}{2\pi^2} g^2 \pi_a + \left(\frac{1}{g^2} -
 (\frac{i\theta}{2\pi^2})^2g^2\right) B_a \right )|_{\d \cal M} =0
\,\,\,\,\,\,\,\,\,\,\,\,\, or
\,\,\,\,\,\,\,\,\,\,\,\,\, 
\delta A_c |_{\d {\cal M}} =0.
\eeq{3.5}
 There are one normal condition (on the left handside (\ref{3.4}))
 and two tangential conditions (on the left handside (\ref{3.5})).
 We will keep these in mind.  We hope to find them
 as boundary constraints required to make 
  the whole  Hamiltonian treatment consistent.
 Let us assume that one can 
 choose a coordinate system such that the normal vector has the form
 $\vec{n}=(1,0,0)$.
 
   We now turn to the Hamiltonian treatment. Using (\ref{3.1}) and
 (\ref{3.3}) one defines the Hamiltonian
\beq
 H = \int\limits_{\Sigma} d^3x \left [ \frac{g^2}{2} \pi_a \pi^a -
 \frac{i\theta}{2\pi^2}g^2\pi_aB^a - \frac{1}{2}\left(\frac{1}{g^2} -
 (\frac{i\theta}{2\pi^2})^2g^2\right) B_a B^a + (\d_a A_0) \pi^a \right ].
\eeq{3.6} 
 Thus one defines the Hamiltonian $H$ and primary 
 constraint $\pi_0$ without using the boundary conditions.
 Introducing the notation
\beq
 \Pi[\Lambda] = \int\limits_{\Sigma} d^3x\,\Lambda(x) \pi^0(x) ,
\eeq{3.7}
 one has the following bracket 
\beq
 \{ \Pi[\Lambda], H\} = {\cal G}[\Lambda] - \int\limits_{\d \Sigma}
  d^2 s \Lambda (n^a \pi_a) ,
\eeq{3.8}
 where ${\cal G}[\Lambda]$ is the Gauss law constraint
\beq
{\cal G}[\Lambda] = \int\limits_{\Sigma} d^3x\,\Lambda(x) \d_a \pi^a(x).
\eeq{3.9}
 Consistency then implies that the right hand of (\ref{3.8})
 must be equal to zero. Thus we are getting the standard Gauss law and
 as well the boundary constraint $n^a\pi_a=0$ (if one assumes
 that $A^0$ is zero on the boundary  there is no boundary 
 term in (\ref{3.8})).

 Following the standard prescription
  one should look at the time evolution of Gauss law (\ref{3.9})
 and the new boundary constraint 
\beq
\Phi_1[\Lambda] = \int\limits_{\Sigma} d^3 x\, \Lambda n^a \pi_a ,
\eeq{3.9a}
 where $\Lambda$ can be thought as test function with support 
 on $\d \Sigma$. The formal computation gives us 
 the following result
\beq
 \{ {\cal G}[\Lambda], H \} = \int\limits_{\d\Sigma} d^2 s\, 
 [ n_b \d_c\Lambda \epsilon^{abc}]
 \left ( \frac{i\theta}{2\pi^2} g^2 \pi_a + \left(\frac{1}{g^2} -
 (\frac{i\theta}{2\pi^2})^2g^2\right) B_a \right ),
\eeq{3.11}
\beq
 \{ \Phi_1[\Lambda], H\} = \int\limits_{\Sigma} d^3 x\, 
 [\d_c(\Lambda n_b)\epsilon^{abc}]
 \left ( \frac{i\theta}{2\pi^2} g^2 \pi_a + \left(\frac{1}{g^2} -
 (\frac{i\theta}{2\pi^2})^2g^2\right) B_a \right ), 
\eeq{3.10}
 where we have used the higher dimensional analog of equation (\ref{a.6a}).
 The bracket (\ref{3.11}) is localized on the boundary and it gives
 us the tangential boundary constraints which exactly coincide with the
  boundary  conditions (\ref{3.5}). The same boundary constraint is given by 
 the bracket (\ref{3.10}) and it is localized on the boundary 
  since $\Lambda$ has support on the boundary only. Therefore we 
 introduce the new boundary constraint
\beq
 \Phi_a[N^a] = \int d^3x\,N_a [\frac{i\theta}{2\pi^2} g^2 
 \pi^a + \left(\frac{1}{g^2} -
 (\frac{i\theta}{2\pi^2})^2g^2\right) B^a ],
\eeq{3.12}
 where it is assumed that $N_a = (0, N_2, N_3)$.
 To decide on the status of boundary constraints $\Phi_1$, $\Phi_2$ and
 $\Phi_3$ one should calculate the following brackets
\beq
 \{ \Phi_1[\Lambda], \Phi_b[N^b]\} = \left(\frac{1}{g^2} -
 (\frac{i\theta}{2\pi^2})^2g^2\right) \int\limits_{\Sigma}
 d^3x\, \Lambda n^a\d^b N^c \epsilon_{abc},
\eeq{3.13}
\beq
\{\Phi_a[N^a], \Phi_b[M^b]\} =  \frac{i\theta}{2\pi^2} g^2
  \left(\frac{1}{g^2} -
 (\frac{i\theta}{2\pi^2})^2g^2\right) \int\limits_{\Sigma}
 d^3x\,\d^a(N^b M^c) \epsilon_{abc}.
\eeq{3.14}
 One notices that the brackets (\ref{3.14}) 
 of $\Phi_a$ ($a=2,3$) are non-zero 
 because of the boundary term on the right hand side of (\ref{3.14}).
 In the bulk such brackets are zero since the constraints 
 $\Phi_a$ are generalization of the chiral condition for 
 two forms \cite{Bengtsson:1997fm}. 
Therefore the brackets of all boundary constraints give a field 
 independent antisymmetric matrix with rank 2. 
 It turns out that there is one first class boundary 
 constraint $n^a \pi_a$ which makes us able to gauge away 
 the normal component of the connection on the boundary. 
 The boundary constraints $\Phi_2$ and $\Phi_3$ are  second
 class constraints which lead to the the following Dirac 
 bracket on the boundary
\beq
 \{ A_2(x), A_3(y)\}|_{\d \Sigma} =  \frac{i\theta}{2\pi^2}
 g^2 \left ( \frac{1}{g^2} + \frac{\theta^2}{4\pi^4} g^2 \right)^{-1}
  \delta^{(2)}(x-y).
\eeq{3.15}
 In analogy with the models considered
 in the previous section we see that  at the boundary the Poisson brackets
 should be replaced by the corresponding Dirac bracket. In general the boundary
 Dirac bracket will depend on the geometry of the boundary. We will discuss this
  elsewhere.   
 The physical interpretation of (\ref{3.15}) is unclear.
  It seems that one will have problems with localizing 
 photons on the boundary.
 Certainly  this subject deserves an
 independent study \cite{} and we do no analyse the 
 boundary theory further here.
 
\section{Discussion and problems}

 In this paper we made an attempt to understand the status of
 the boundary conditions within the Hamiltonian formalism motivated
  by the quantum theory. 
 We have shown  that boundary conditions 
 can arise in a purely algebraic fashion 
 as Hamiltonian boundary constraints. Their existence
 is necessary to make the whole Hamiltonian treatment consistent. 
 Our arguments were based on four examples: the topological sigma model,
  the open string theory
 with and without a B-field and electrodynamics with a topological term.
 For some systems it is important to motivate that the boundary 
 conditions can be treated as Hamiltonian constraints. This type
 of systems has non trivial boundary conditions which mix   
 momenta and coordinates. Such boundary conditions change 
 the canonical brackets on the boundary drastically 
and therefore they are very
 important for the quantization of the system as whole. However as
 we saw in some instances (e.g.,(\ref{1.8}))
  problems can arise with the definition 
 of the Dirac bracket on the boundary. 
 To resolve those problems one needs more
 insight into the models. In other cases there is no ambiguity 
 in defining the modified symplectic structure
 (e.g., (\ref{a.7}), local part of (\ref{2.9}) and (\ref{3.15})). 

 As can be seen from the last example  the boundary conditions
 give rise not only to second class constraints but also
 to first class constraint. It is unclear how this kind of
 boundary constraint should be applied in the quantum theory. 
 We hope to return to this question and do  explicit 
 calculations for this model in the presence of a simple boundary. 

 From a technical point of view the present approach is based
 on rules for dealing with the following expression
\beq
 \int\limits_{\Sigma} d^d x  \int\limits_{\Sigma} d^d x' f(x) g(x') 
 D_x D_{x'} \delta(x-x')
\eeq{D1}
 where $D_x$ ($D_{x'}$) is some differential operator. As was pointed out 
 in \cite{Solovev:1993zf} the expression (\ref{D1}) is not in general well defined
  on a closed domain. In the present paper we considered 
 simple models with a few derivatives and therefore it was no problem 
 to define (\ref{D1}) in a reasonable way. 
 However in general one should address this question more carefully. 

 It would be interesting to study an interacting theory
 like Yang-Mills theory and gravity systems in this context.
  We hope to treat
 these questions elsewhere.

\bigskip

\begin{flushleft} {\Large\bf Acknowledgments} \end{flushleft}

 It is pleasure to thank Ingemar Bengtsson, who
 has promoted my interest in the subject and who has helped
 a lot during the preparation of this work. I am grateful to
 Ingermar Bengtsson and Ulf Lindstr\"om for 
 reading and commenting on the manuscript. I thank M.M.Sheikh-Jabbari
 and  V.O.Soloviev for bringing the relevant 
 references to my attention.

\end{document}